\documentclass[aps,pre,twocolumn,superscriptaddress,showpacs]{revtex4-1}
\usepackage{graphicx}
\usepackage{dcolumn}
\usepackage{bm}
\usepackage{epsfig}
\usepackage{color}

\begin{document}


\title{Understanding the Local Flow Rate Peak of a Hopper Discharging
  Discs through an Obstacle Using a Tetris-like Model}


\author{Guo-Jie Jason Gao}
\email{koh.kokketsu@shizuoka.ac.jp, gjjgao@gmail.com}
\affiliation{Department of Mathematical and Systems Engineering,
  Shizuoka University, Hamamatsu, Shizuoka 432-8561, Japan}

\author{Jerzy Blawzdziewicz} \affiliation{Department of Physics, Texas
  Tech University, Lubbock, TX 79409-1051, USA}
\affiliation{Department of Mechanical Engineering, Texas Tech
  University, Lubbock, TX 79409-1021, USA}

\author{Michael C. Holcomb} \affiliation{Department of Physics, Texas
  Tech University, Lubbock, TX 79409-1051, USA}

\author{Shigenobu Ogata} \affiliation{Department of Mechanical Science
  and Bioengineering, Osaka University, Toyonaka, Osaka 560-8531,
  Japan} \affiliation{Center for Elements Strategy Initiative for
  Structural Materials (ESISM), Kyoto University, Sakyo, Kyoto
  606-8501, Japan}

\date{\today}

\begin{abstract}
Placing a round obstacle above the orifice of a flat hopper
discharging uniform frictional discs has been experimentally and
numerically shown in the literature to create a local peak in the
gravity-driven hopper flow rate. Using frictionless molecular dynamics
(MD) simulations, we show that the local peak is unrelated to the
interparticle friction, the particle dispersity, and the obstacle
geometry. We then construct a probabilistic Tetris-like model, where
particles update their positions according to prescribed rules rather
than in response to forces, and show that Newtonian dynamics are also
not responsible for the local peak. Finally, we propose that the local
peak is caused by an interplay between the flow rate around the
obstacle, greater than the maximum when the hopper contains no
obstacle, and a slow response time, allowing the overflowing particles
to converge well upon reaching the hopper orifice.
\end{abstract}


\maketitle


\section{Introduction}
\label{introduction}
Placing a round obstacle near the orifice of a granular hopper has
recently been shown to reduce particle clogging \cite{zuriguel11,
  zuriguel14, katsuragi18} and locally speed up the gravity-driven
hopper flow rate \cite{zuriguel11, lozano12, alonso-marroquin12,
  zuriguel15, alonso-marroquin16}. The flow rate exhibits a peak as
the obstacle is placed an optimal distance away from the orifice of
the hopper.  Intrigued by these findings, we want to understand what
role interparticle friction, particle size dispersity, obstacle
geometry, and the Newtonian dynamics that produce interparticle
cooperative motion plays in the appearance of a flow rate peak.

To study the role of interparticle friction, we use a molecular
dynamics (MD) method to simulate frictionless, monodisperse discs
flowing about a round obstacle placed near the orifice of a hopper.
We measure the gravity-driven flow rate $J_a$ in terms of number of
discs out of the hopper per unit time. Interestingly, eliminating
friction in the system does not prevent the local flow rate peak from
occurring even though the peak value, normalized by the flow rate
$J_o$ when the hopper contains no obstacle, becomes smaller than
unity. Changing particle dispersity from monodisperse to bidisperse or
altering the obstacle shape from round to nearly flat also does not
annihilate the local peak. The MD results suggest that interparticle
friction, particle dispersity, and obstacle geometry are not
fundamental factors responsible for the hopper flow rate peak. We
therefore propose a necessary condition for predicting the occurrence
of the flow rate peak: the peak should appear soon after the flow rate
$J_i$ measured at the obstacle becomes greater than $J_o$. This
condition approximately forecasts where the local flow rate peak
happens without resorting to the continuum theory of granular hopper
flow. We successfully verify the proposed necessary condition,
$J_i/J_o>1$, using frictionless MD simulations.

In light of our MD results, we further reduce the dynamics of the
system by completely switching off Newton's equations of motion
through the introduction of a Tetris-like model. Circular particles of
equal size in a hopper update their positions sequentially under a set
of prescribed rules, similar to the classical video game
\textit{Tetris}, forming a probability-driven hopper flow with
adjustable driving strength. Another identically-named but more
simplified model has been used to study the compaction of granular
materials under vibration \cite{nicodemi97}. In the Tetris-like model,
particles interact with their nearest neighbors by means of
trial-and-error position-update cycles. The model maintains the
essential dynamics of granular materials through non-overlap
geometrical constraints, which occasionally creates particle
clogging. The lack of Newtonian dynamics in the Tetris-like model
inherently suspends interparticle cooperative motion via
forces. Surprisingly, we can observe the local flow rate peak so long
as the necessary condition $J_i/J_o>1$ is satisfied and a slow
response time explained below even in this probabilistic model with
minimal dynamics, and the maximum value of the normalized local flow
rate peak $J_a/J_o$ can be below or above unity, similar to the
results reported in studies using inanimate frictionless or frictional
objects, or even animals \cite{zuriguel15_1, zuriguel16}.

Based on the results of the Tetris-like model, we suggest that the
local flow rate peak can be qualitatively understood by combining a
linearly increasing $J_i$, satisfying the necessary condition
$J_i/J_o>1$, and a slow response time, restricting the flow rate of
particles passing the obstacle to reaching the hopper orifice.  The
slow response time allows discharged particles to merge more fully as
they move towards the hopper orifice, resulting in the merged
particles having a packing density high enough to show a local peak of
$J_a/J_o$ as long as the obstacle is placed at some optimal
height. We expect that a slower response time corresponds to a larger
variation in the number of times a particle fails to update its
position, $n_{hit}$, as a function of obstacle location.  We observe
the expected relation in representative cases showing local flow rate
peaks, and verify our idea successfully.

Below we elaborate on our Tetris-like model which generates the
probability-driven hopper flow in section \ref{tetris model}, followed
by quantitative investigation of the hopper flow rates in section
\ref{results_and_discussions}. Finally, we conclude our study in
section \ref{conclusions}. Since we focus on the Tetris-like model in
the main text, we leave the description of our frictionless MD
simulations to the appendix section \ref{MD method} for reference.

\section{The Tetris-like model}
\label{tetris model}
\begin{figure}
\includegraphics[width=0.49\textwidth]{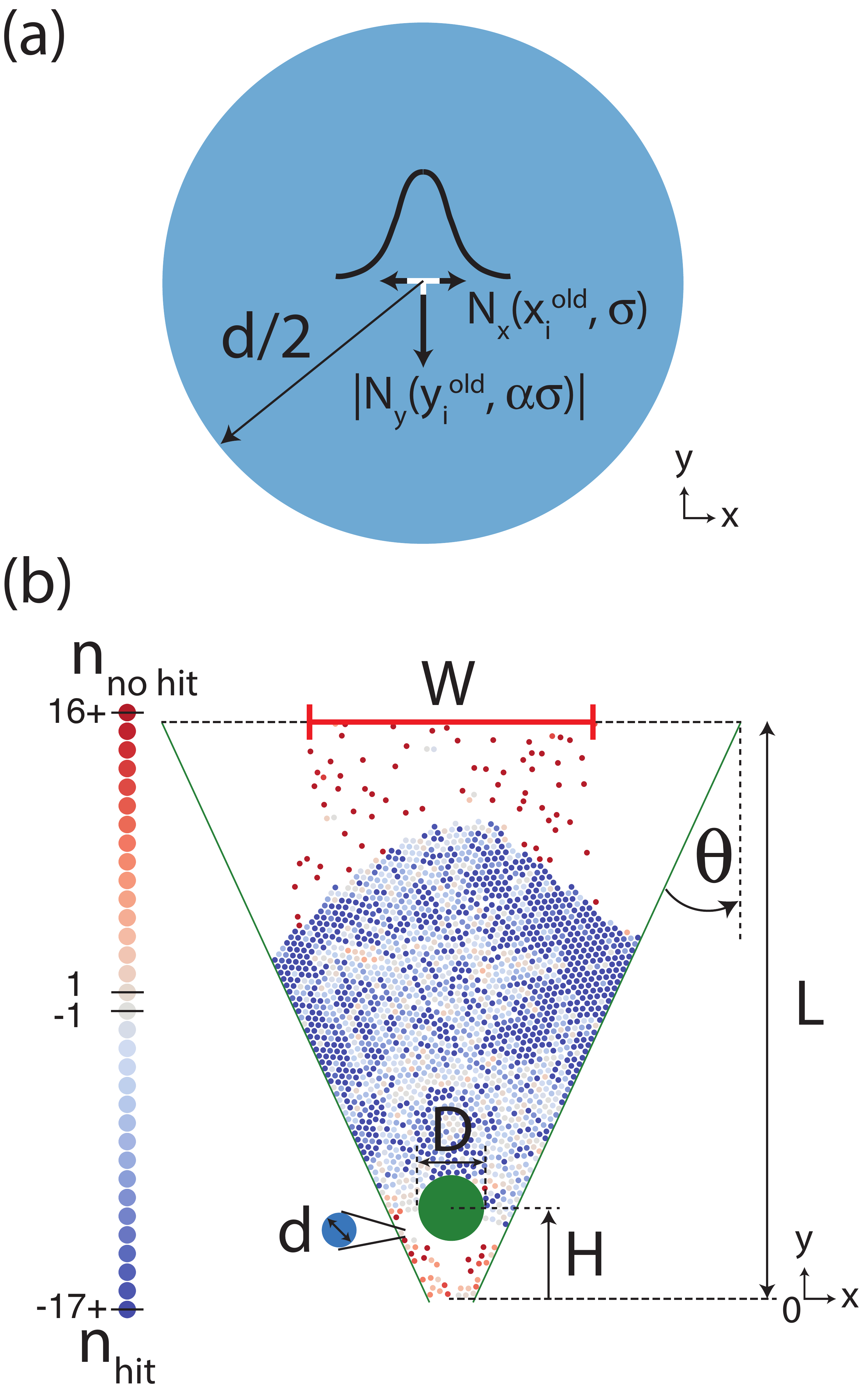}
\caption{\label{fig:tetris_model} (Color online) (a) The objects of
  the Tetris-like model are demonstrated by the disc particle of
  diameter $d$ (blue circle).  The particle's next position is
  governed by two independent Gaussian functions, $N_x$ and $N_y$
  having (mean, standard deviation) equal to ($x_i^{old},\sigma$) and
  ($y_i^{old},\alpha \sigma$), respectively. $\sigma = 0.05d$ and
  $\alpha$ is a variable of the driving strength in this study. (b)
  The simulation setup of a symmetric hopper (green lines) with equal
  height and top-width $L$, and a hopper angle $\theta$. An obstacle
  (green circle) of diameter $D$ is located at a height $H$ above the
  hopper orifice. The hopper discharges particles, colored by $n_{no
    \ hit}$ or $n_{hit}$ recording the history of successfully or
  unsuccessfully updating their positions successively in a linear
  scale. The discharged particles reenter the hopper from its top
  border within $W = 0.5L$ through a random dispersion. The snapshot
  is taken with $\alpha=0.128$.}
\end{figure}

To study the externally-driven granular hopper flow without the
involvement of Newtonian dynamics, we propose a purely geometrical
Tetris-like model.  Tetris is a classic video game where solid objects
with given shapes drop down one by one towards the bottom of the
playing field. The player can shift the objects horizontally at will
during the dropping process. In our Tetris-like model, each circular
particle $i$ of equal diameter $d$ has exactly one chance per
position-update cycle to change its horizontal ($x$) and vertical
($y$) positions from $(x_i^{old}, y_i^{old})$ to $(x_i^{new},
y_i^{new})$ , according to
\begin{equation} \label{eqn_x}
x_i^{new} = {N_x}(x_i^{old},\sigma),
\end{equation}
and
\begin{equation} \label{eqn_y}
y_i^{new} = \left| {{N_y}(y_i^{old},\alpha \sigma)} \right|,
\end{equation}
where $N_x$ and $N_y$ are two independent Gaussian functions. $N_x$
has a mean at $x_i^{old}$ and a standard deviation $\sigma$, while
$N_y$ has a mean at $y_i^{old}$ and a standard deviation
$\alpha\sigma$, as shown schematically in
Fig.\ref{fig:tetris_model}(a). The absolute value about $N_y$ forbids
backward movements of particles. We choose $\sigma=0.05d$ throughout
this study, and $\alpha$ is a control parameter representing the
strength of driving particles towards the orifice of the hopper,
similar to the driving force of discharging animate or inanimate
particles through constrictions \cite{zuriguel15}.

There are $N=2048$ randomly placed particles in the hopper in the
beginning of a simulation, and we update their positions sequentially
using a random list, renewed repetitively per position-update
cycle. The Tetris-like model accepts a position update of a particle
if it creates no overlap with any other objects in the system. A
positive parameter $n_{no\ hit}$, starting from zero, is increased by
one to save this successful trial. Otherwise, the update is rejected
and the particle stays still. Similarly, a negative parameter
$n_{hit}$, also starting from zero, is decreased by one to save this
failed trial. $n_{no\ hit}$ is reset when $n_{hit}$ becomes nonzero,
and vice versa. The hopper is geometrically symmetric and has a height
$L=83d$ and a hopper angle $\theta=0.4325$ radians. A circular
obstacle of diameter $D=0.112L$ and $D/d=9.296$ is placed along the
symmetric axis of the hopper a height $H$ above its orifice. To
maintain a constant $N$, a particle leaving the hopper will reenter it
from its top border with the particle's $x$ position randomly
reassigned within a range $W \in [-L/4,L/4]$. A snapshot of the
probability-driven hopper flow is shown in
Fig.\ref{fig:tetris_model}(b). Further details about the Tetris-like
model can be found in our previous study \cite{gao18}.

\begin{figure}
\includegraphics[width=0.36\textwidth]{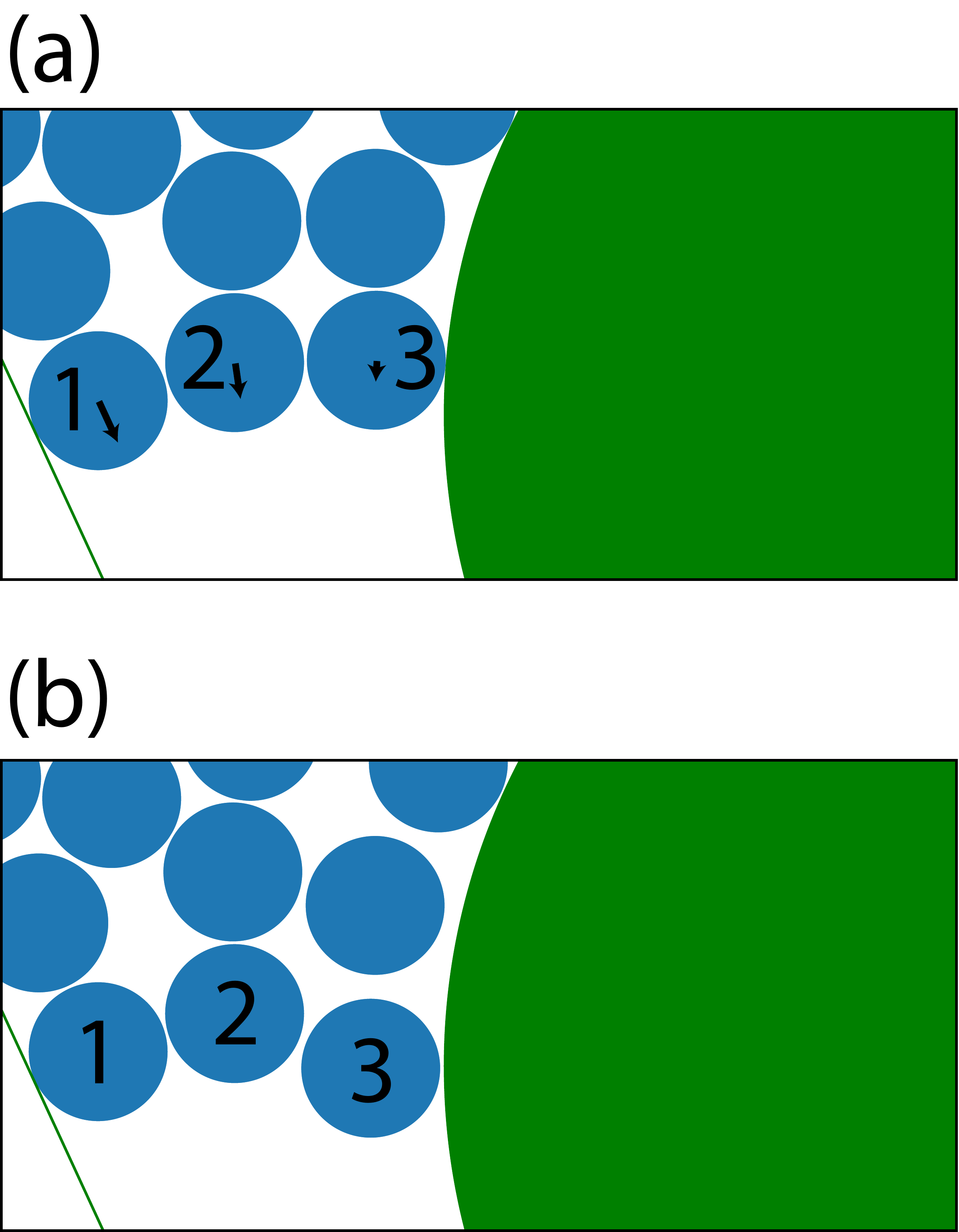}
\caption{\label{fig:arching} (Color online) Snapshots of an arch
  forming and breaking event in the Tetris-like model with
  $\alpha=1.0$. (a) Particles 1, 2, and 3 form the arch.  The arrows
  on the particles indicate the minimal jump distance for escaping
  from the geometrical constraint without creating any overlap with
  their neighboring particles. (b) The arch breaks when particle 3,
  having the shortest jump distance for escaping and therefore the
  highest chance of success, is the first to undermine the arch.}
\end{figure}

Using the Tetris-like model, we measure the actual flow rate $J_a$ in
terms of the average number of particles passing the hopper orifice
per position-update cycle, and define $J_o$ as the value of $J_a$
while the hopper contains no obstacle.  We also measure the average
number of particles $J_i$ flowing out of the two channels between the
obstacle and the hopper walls, essentially considering the flow rate
of an imperfect hopper with the part of its orifice lower than the
center of the obstacle removed.  Our Tetris-like model sometimes
encounters persistent clogging events due to geometrical particle
arching, as shown by the explanatory snapshots in
Fig.\ref{fig:arching}. These events exist within a completely
different timescale. This issue becomes serious when the driving
strength $\alpha$ is very weak, similar to extremely slow grain
velocities cause orders of magnitude higher hopper clogging
probability found in experiments \cite{zuriguel18}. To ensure that
measured flow rates are free from persistent clogging events we
discard simulation data containing clogging events lasting longer than
$10,000$ position-update cycles, a practice similar to using vibration
to resume the clogged hopper flow in experiments \cite{zuriguel15}.

\section{Results and Discussions}
\label{results_and_discussions}
Below, we show the normalized hopper flow rates $J_a/J_o$ as a function
of the driving strength $\alpha$. We then focus on two exemplary
cases: $\alpha=0.333$ showing a local flow rate peak and
$\alpha=0.439$ showing no peak. Finally, we offer a plausible
mechanism for the observed peaks backed by our simulation evidence.

\subsection{The effect of the driving strength $\alpha$ on the hopper flow rate}
\label{alpha_dependence}
\begin{figure}
\includegraphics[width=0.45\textwidth]{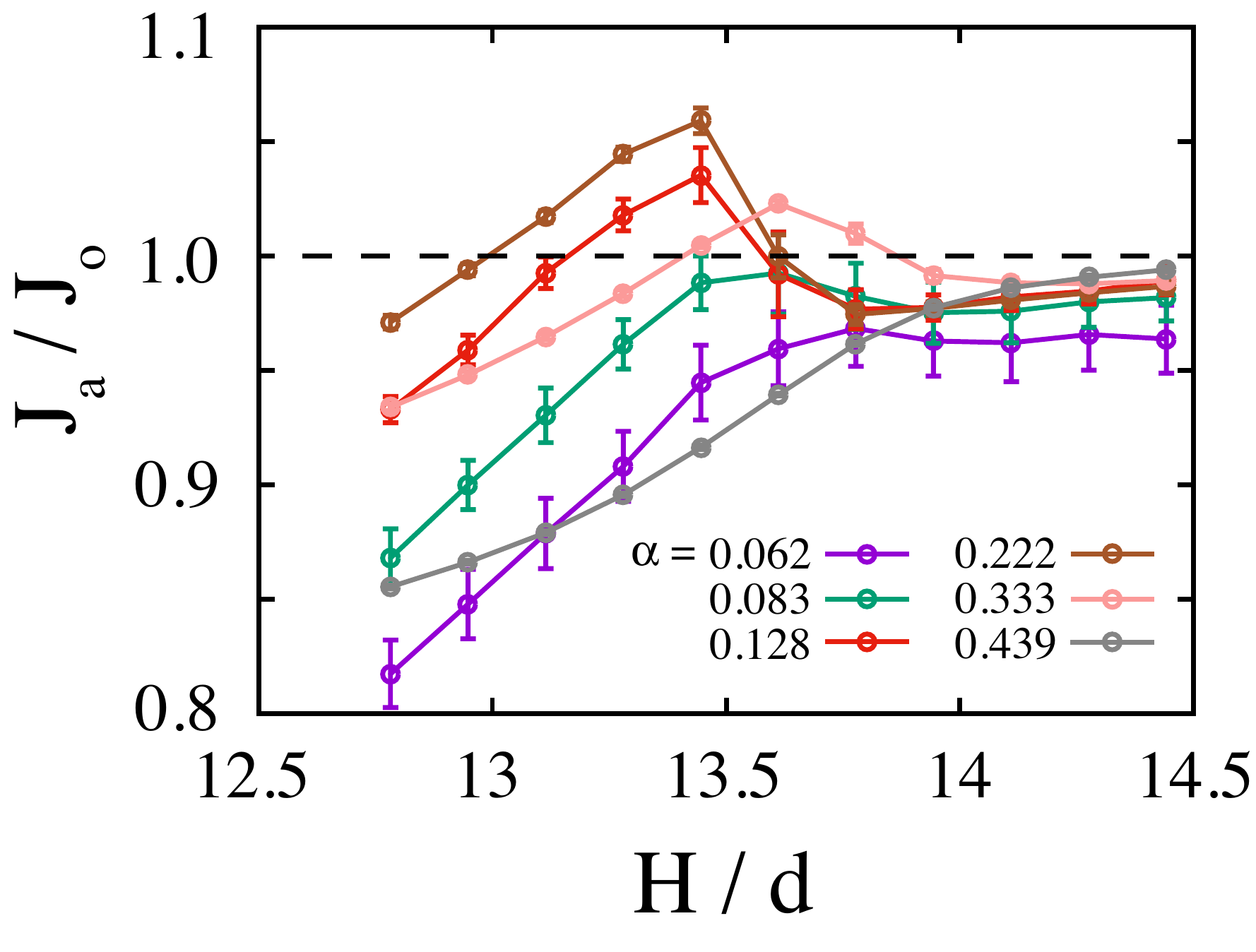}
\caption{\label{fig:flowrate_py_all} (Color online) Hopper flow rates
  $J_a$ measured at the hopper orifice with $\alpha=0.062$ (purple),
  $0.083$ (green), $0.128$ (red), $0.222$ (brown), $0.333$ (pink), and
  $0.439$ (grey), respectively. The error bars of each curve are
  obtained using $45$ different initial conditions.}
\end{figure}

To test the effect of the driving strength $\alpha$ on the normalized
hopper flow rate $J_a/J_o$, we tried six different values of $\alpha$
between $0.062$ and $0.439$ and measured the corresponding
$J_a/J_o$. The results are shown in Fig.\ref{fig:flowrate_py_all}. We
can see that when the driving strength is weak ($\alpha=0.062$ and
$0.083$), $J_a/J_o$ exhibits a mild local peak below unity as the
obstacle is placed around $H/d=13.5$. A similar phenomenon has been
found using frictionless MD simulations, shown in
Fig. \ref{fig:hopper_flow_rate_comparison} in the appendix section
\ref{MD simulation results}. When we increase $\alpha$ to $0.128$ and
$0.222$, the local peak value of $J_a/J_o$ increases to greater than
unity; the peak value of $J_a/J_o$ decreases slightly as $\alpha$
becomes $0.333$. Similar enhanced flow rates have been reported using
frictional MD simulations in the literature
\cite{alonso-marroquin12}. Finally, when we increase $\alpha$ to
$0.439$, the local peak disappears and $J_a/J_o$ becomes a
monotonically increasing function of $H/d$, consistent with the
findings of another experimental study \cite{katsuragi17}.

\subsection{Examining the hopper flow rate with or without a local peak}
\label{examination}
We take a closer look at two representative cases to learn more about
what happens when a local peak in $J_a/J_o$ is shown or not shown:
$\alpha=0.333$ and $0.439$, respectively.

\begin{figure}
\includegraphics[width=0.45\textwidth]{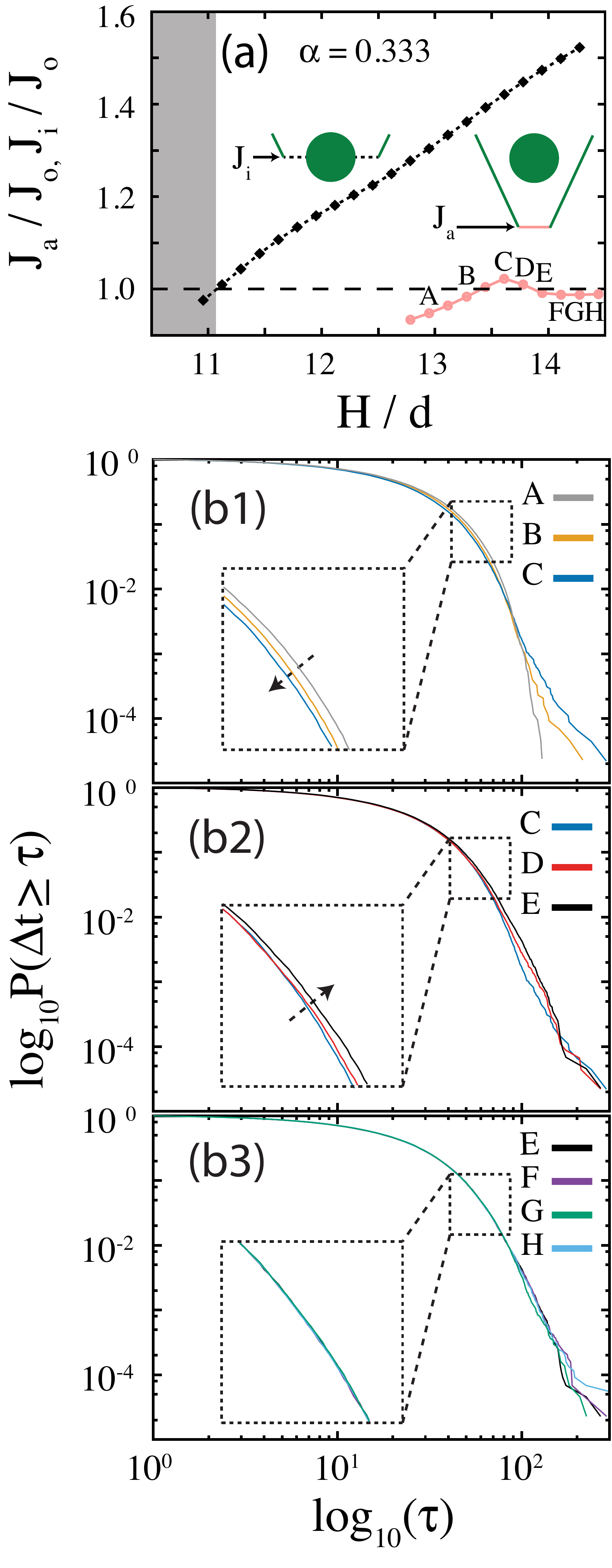}
\caption{\label{fig:flowrate_py_0.143} (Color online) (a) Hopper flow
  rates $J_i$ measured at the obstacle height (black dotted line with
  diamonds) and $J_a$ measured at the hopper orifice (pink solid line
  with circles) with $\alpha=0.333$. The error bars of $J_i$ and $J_a$
  are obtained using $13$ and $45$ different initial conditions,
  respectively. The fluidized flow regime, where $J_i<J_o$, is
  shaded. (b1), (b2) and (b3) show representative $P(\Delta t \ge
  \tau)$, zoomed in the dashed boxes, using log-10 scales for both
  axes within three continuous ranges, A-B-C, C-D-E, and E-F-G-H, on
  the $J_a$ curve.}
\end{figure}

In Fig. \ref{fig:flowrate_py_0.143}(a), we plot $J_a/J_o$, the
normalized flow rate leaving the hopper orifice, and $J_i/J_o$, the
normalized flow rate measured at the obstacle, as a function of $H/d$
when $\alpha=0.333$. We can see clearly that $J_a/J_o$ exhibits a
local peak around $H/d=13.6$ after $J_i/J_o$ becomes greater than
unity around $H/d=11.1$, which shows that $J_i/J_o>1$ is a necessary
condition for a local flow rate peak of $J_a/J_o$.

We then calculate the time lapse $\Delta t$ between the egress of two
consecutive particles leaving the hopper at position-update cycles $i$
and $i+\Delta t$. From the $\Delta t$ data, we plot the complementary
cumulative distribution function $P(\Delta t \ge \tau)$, which gives
the probability of finding a time lapse $\Delta t$ equal to or larger
than $\tau$ \cite{zuriguel14, zuriguel15}. For a given value of $H/d$,
we build $P(\Delta t \ge \tau)$ using $990,000$ position-update
cycles. It is possible for two or more particles to leave the hopper
during the same position-update cycle, but we treat this as a single
egress event and obtain only one $\Delta t$ from it. The
multi-particle egress is rare and accounts for less than $1\%$ of the
total egress events in the data reported here.

Fig. \ref{fig:flowrate_py_0.143}(b1-b3) show $P(\Delta t \ge \tau)$ in
three continuous ranges of $H/d$, where $J_a/J_o$ first increases
(Range A-B-C), then decreases (Range C-D-E), and finally reaches a
steady value (Range E-F-G-H) under the mild driving strength
$\alpha=0.333$. In Range A-B-C with a moderate $J_i/J_o>1$, the mild
driving strength allows the two groups of particles, discharged from
the two sides of the obstacle toward the center of the hopper orifice
through a narrowing passage, to merge more efficiently by means of
non-overlapping particle position-updates without hitting the hopper
walls too often. As a result, we observe an increasing $J_a/J_o$, and
upon reaching the hopper orifice, the concentration of confluent
particles can be high enough to deliver a $J_a/J_o>1$ if we place the
obstacle at the optimal height C, as shown by the A-B-C series of
reducing $P(\Delta t \ge \tau)$ in
Fig. \ref{fig:flowrate_py_0.143}(b1). However, in Range C-D-E with a
stronger $J_i/J_o>1$, the clogging effect appears, shown by the C-D-E
series of increasing $P(\Delta t \ge \tau)$ in
Fig. \ref{fig:flowrate_py_0.143}(b2). The overall effect is a
decreasing $J_a/J_o$. Lastly, in Range E-F-G-H, because the obstacle
is located far from the hopper orifice, its influence on the flow rate
becomes negligible, and the E-F-G-H series of $P(\Delta t \ge \tau)$
shows no clear trend, as can be seen in
Fig. \ref{fig:flowrate_py_0.143}(b3).

\begin{figure}
\includegraphics[width=0.45\textwidth]{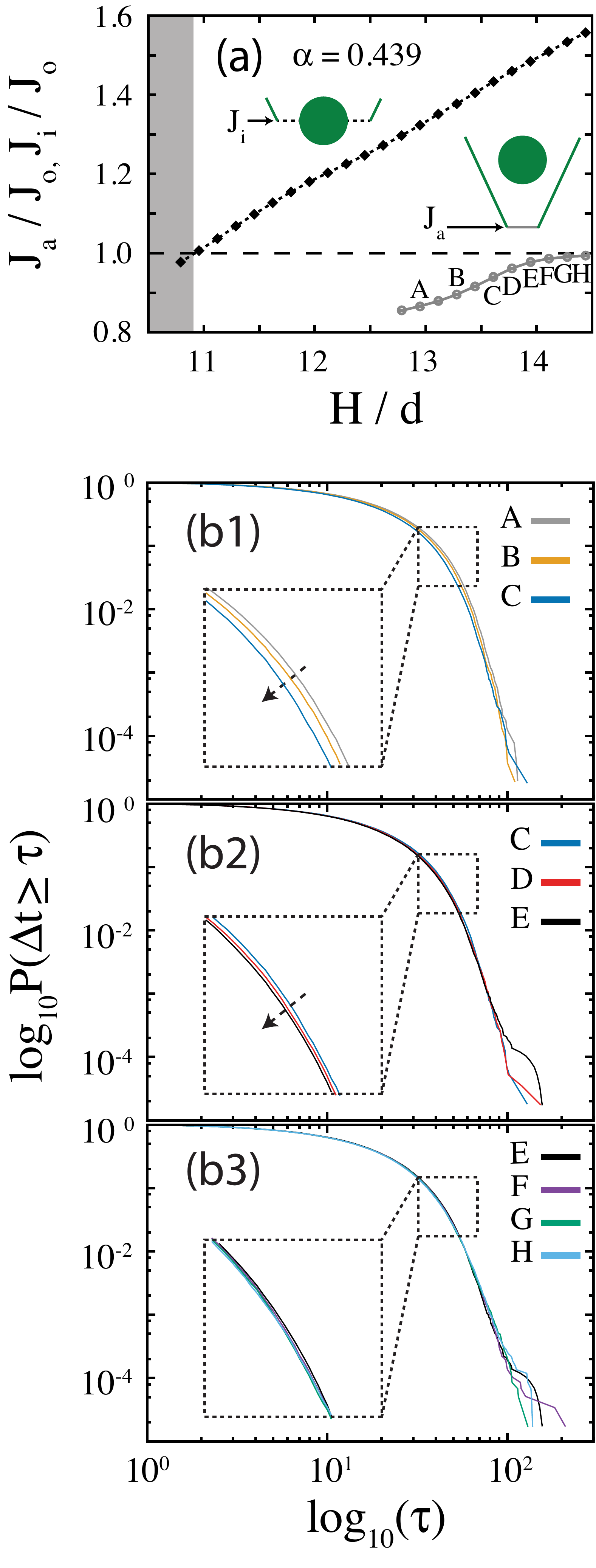}
\caption{\label{fig:flowrate_py_0.180} (Color online) The same plots
  as Fig. \ref{fig:flowrate_py_0.143}, except $\alpha=0.439$.}
\end{figure}

Similarly, in Fig. \ref{fig:flowrate_py_0.180}(a), we plot $J_a/J_o$
which exhibits no local peak under the strong driving strength
$\alpha=0.439$, and the corresponding $J_i/J_o$ as a function of
$H/d$. Fig. \ref{fig:flowrate_py_0.180}(b1-b3) show the related
$P(\Delta t \ge \tau)$ in the same three ranges of $H/d$ as before.
Unlike the milder driving strength $\alpha=0.333$, where $J_a/J_o$ can
be greater than unity, discharged particles prompted by the stronger
driving strength $\alpha=0.439$ block one another and hit the hopper
walls more often, which on average gives a higher $n_{hit}$ per
particle, resulting in less efficient merging on the way toward the
hopper orifice and stagnation in the Tetris-like hopper.  We can still
observe an increasing $J_a/J_o$, as shown by the A-B-C series of
reducing $P(\Delta t \ge \tau)$ in
Fig. \ref{fig:flowrate_py_0.180}(b1); however, the concentration of
confluent particles can never deliver a $J_a/J_o>1$. Due to $J_a/J_o$
remaining lower than unity, in the following range we do not observe a
substantial counteractive clogging effect as in
Fig. \ref{fig:flowrate_py_0.143}(b2), and $P(\Delta t \ge \tau)$ keeps
decreasing until it reaches a steady value, as shown in
Fig. \ref{fig:flowrate_py_0.180} (b2) and (b3), respectively.

\subsection{An explanation for the local flow rate peak}
\label{idea_model}
In our previous work \cite{gao18}, we suggest that ${J_a} \sim
{{{J_i}} \mathord{\left/ {\vphantom {{{J_i}} {{P_c}}}} \right.
    \kern-\nulldelimiterspace} {{P_c}}}$, where $P_c$ is the
probability of particle clogging below the obstacle. Reasonably, the
flow rate leaving the hopper $J_a$ should be proportional to $J_i$,
measuring the number of particles released from the two passages
between the obstacle and the hopper walls. As $J_i$ increases with
higher placement of the obstacle, more particles comes out from the
passages. As a result, the clogging probability $P_c$ should also
become higher, a reasoning consistent with studies claiming that the
probability for a given particle to be able to participate in a clog
is constant in granular hopper flow \cite{durian13,
  durian15}. Besides, we also assume that the increasing $P_c$ puts a
strong constraint on the value of $J_a$.

\begin{figure}
\includegraphics[width=0.48\textwidth]{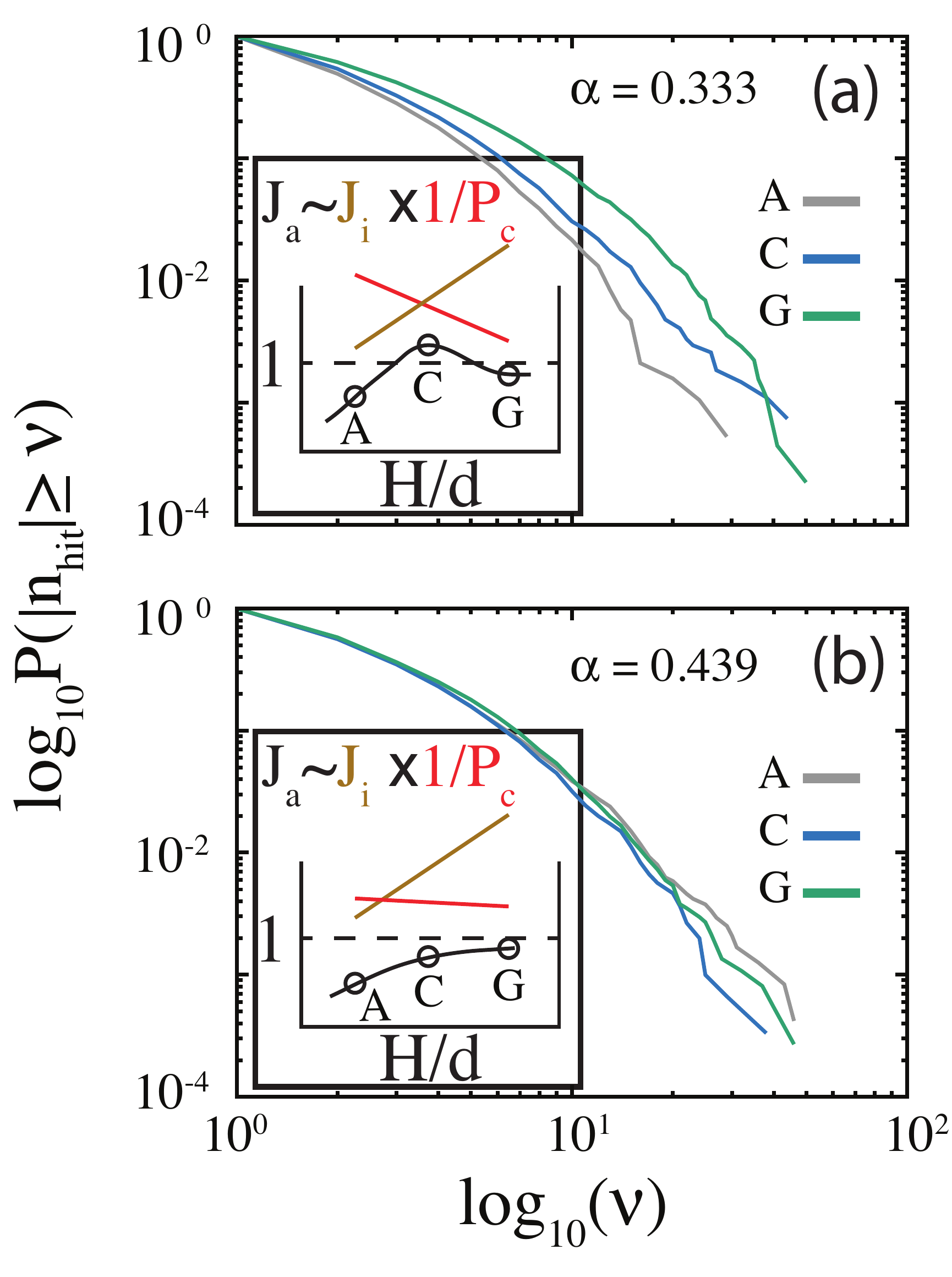}
\caption{\label{fig:surv_func_N_hit} (Color online) Representative
  $P(\left| {{n_{hit}}} \right| \ge \nu )$, using log-10 scales for
  both axes, across a range of A-C-G on the $J_a$ curves of (a)
  $\alpha=0.333$ and (b) $\alpha=0.439$ as in
  Fig. \ref{fig:flowrate_py_0.143}(a) and
  Fig. \ref{fig:flowrate_py_0.180}(a), respectively. The insets show
  schematically drawn ${J_a} \sim {J_i} \times \frac{1}{{{P_c}}}$,
  where $J_i$ (brown) are alike, but $1/P_c$ (red) is steeper in (a),
  where the driving strength $\alpha$ is weaker, and allows a local
  peak of the hopper flow $J_a$.}
\end{figure}

In this study, we explicitly show in
Fig. \ref{fig:flowrate_py_0.143}(a) and
Fig. \ref{fig:flowrate_py_0.180}(a) that $J_i$ is a linear function of
$H/d$ obtained by discarding data of $J_i$ containing clogging events
in the Tetris-like model, as discussed in section \ref{tetris model}
and shown in Fig.\ref{fig:arching}. In addition, we assume that $P_c$
is proportional to the absolute value of $n_{hit}$, defined as a
negative number recording how many times a particle fails to update
its position due to creating an overlap with other objects in the
hopper, that is, ${P_c} \sim \left| {{n_{hit}}} \right|$. Using only
the $n_{hit}$ data of particles whose positions are below the
obstacle, we build its complementary cumulative distribution function
$P(\left| {{n_{hit}}} \right| \ge \nu )$, which gives the probability
of finding a particle failing to move equal to or larger than $\nu$
times. The results of $P(\left| {{n_{hit}}} \right| \ge \nu )$ for
$\alpha=0.333$ and $\alpha=0.439$, covering the same three ranges of
$H/d$ discussed in Fig. \ref{fig:flowrate_py_0.143} and
Fig. \ref{fig:flowrate_py_0.180}, are shown in
Fig. \ref{fig:surv_func_N_hit}.

In Fig. \ref{fig:surv_func_N_hit}(a) with the weaker driving strength
$\alpha=0.333$, we can see that the variation of $P(\left| {{n_{hit}}}
\right| \ge \nu )$ is larger.  This indicates a slower response time
of the system, once it senses an increasing supply of particles $J_i$
in the upper stream and tries to regulate the output flow rate $J_a$
by increasing $n_{hit}$. In the inset, we schematically draw ${J_a}
\sim {J_i} \times \frac{1}{{{P_c}}}$. A linearly increasing $J_i$ and
a steep ${1 \mathord{\left/ {\vphantom {1 {{P_c}}}} \right.
    \kern-\nulldelimiterspace} {{P_c}}} \sim {1 \mathord{\left/
    {\vphantom {1 {\left| {{n_{hit}}} \right|}}} \right.
    \kern-\nulldelimiterspace} {\left| {{n_{hit}}} \right|}}$
presumably can allow $J_a$ to overshoot briefly and exhibit a local
peak, as observed in our simulation results and reported in other
experimental and numerical studies. In the literature
\cite{zuriguel11}, it has been claimed that an effective pressure
reduction in the region of arch formation below the obstacle and above
the orifice can explain the local hopper flow rate peak. However, in
the Tetris-like model where force and pressure are undefined, we can
still see the local flow rate peak; our explanation offers a novel
point of view.

On the other hand, if the driving strength is stronger, the response
time of the system becomes faster and, therefore, the variation in
$P(\left| {{n_{hit}}} \right| \ge \nu )$ is smaller, as shown in
Fig. \ref{fig:surv_func_N_hit}(b). In its inset, the nearly flat ${1
  \mathord{\left/ {\vphantom {1 {{P_c}}}} \right.
    \kern-\nulldelimiterspace} {{P_c}}} \sim {1 \mathord{\left/
    {\vphantom {1 {\left| {{n_{hit}}} \right|}}} \right.
    \kern-\nulldelimiterspace} {\left| {{n_{hit}}} \right|}}$
restricts $J_a$ from having a local peak. We will pursue the
functional form of ${J_a} \sim {{{J_i}} \mathord{\left/ {\vphantom
      {{{J_i}} {\left| {{n_{hit}}} \right|}}} \right.
    \kern-\nulldelimiterspace} {\left| {{n_{hit}}} \right|}}$ in
future work.

\section{Conclusions}
\label{conclusions}
Using frictionless MD simulations, we show that the interparticle
friction, obstacle geometry, and particle dispersity have no
fundamental contribution to the occurrence of a local peak in the
actual hopper flow rate $J_a$, as recently reported in frictional
systems with discs passing about a round obstacle
\cite{alonso-marroquin12}. Guided by our frictionless MD results, we
suggest a necessary condition, $J_i/J_o>1$, for observing the local
peak, formulated in terms of the flow rate $J_i$ measured at the
obstacle and the maximum flow rate $J_o$ at the orifice when the
hopper contains no obstacle. Our evidence from frictionless MD
simulations supports the proposed necessary condition well.

While the necessary condition identifies when one can expect a local
flow rate peak to happen, it does not explain the reason behind
it. The local effect is still perplexed by factors such as the
interparticle collaborative motion that emerges from the Newtonian
dynamics.  For example, a group of particles can crystallize above the
obstacle or there could be coordinated motions between particles below
the obstacle before they leave the hopper. To reveal the fundamental
cause for the focused phenomenon we proposed a Tetris-like model to
reduce the dynamics of the system to its bare-bones minimum.  In this
model, particles moved sequentially according to prescribed
probability functions without any communication through Newton's
equations of motion, creating an artificial probability-driven hopper
flow.  The non-overlap position-update procedure in the model allows
particles to clog. Strikingly, the peak of $J_a$ still occurs. This
serves as indisputable evidence that the Newtonian dynamics and
associated interparticle collaborative motion are not essential for
this local phenomenon.

Enlightened by the results of our Tetris-like model, we devise a
mechanism to explain the local flow rate peak by introducing a
response time that uses detection of $J_i/J_o>1$ to restrict flow rate
$J_a$ exiting the hopper. The mechanism utilizes the dependence of
$J_a$ on the linearly increasing flow rate $J_i$ and the effect of the
hopper below the obstacle.  As $J_a$ is regulated within the response
time, the particles below the obstacle rearrange themselves, subject
to the non-overlap condition, as they move toward the hopper
orifice. If the necessary condition $J_i/J_o>1$ is satisfied and the
response time is slower, the two groups of particles discharged from
the two passages between the obstacle and the hopper wall can merge
better.  This promotes a higher packing density at the hopper orifice,
allowing the local peak of $J_a/J_o$ to be possible. We link a slower
response time with a larger variation in the clogging probability
$P_c$ within the space between the obstacle and the hopper
orifice. $P_c$ is quantified by measuring the number of times a
particle fails to update its position, $n_{hit}$, as a function of the
obstacle location.  For representative cases where $J_a/J_o$ exhibits
a local peak, we do observe the expected large variations in
$n_{hit}$; however, for cases showing no local peak of $J_a/J_o$, the
variations become reasonably small. In this, we indirectly verify the
proposed mechanism. Our Tetris-like model offers an example of how one
can elucidate underlying mechanisms of a complicated local phenomenon
in an athermal granular system using a simplified model that preserves
only the essential dynamics.

\section{acknowledgments}
GJG gratefully acknowledges financial support from startup funding of
Shizuoka University (Japan).

\section{Compliance with ethical standards}
Conflict of Interest: The authors declare that they have no conflict
of interest. The research presented did not involve human participants
and/or animals.

\section{Appendix: Frictionless MD simulations}
\label{MD method}

\subsection{System geometry}
\label{MD System geometry}
\begin{figure}
\includegraphics[width=0.45\textwidth]{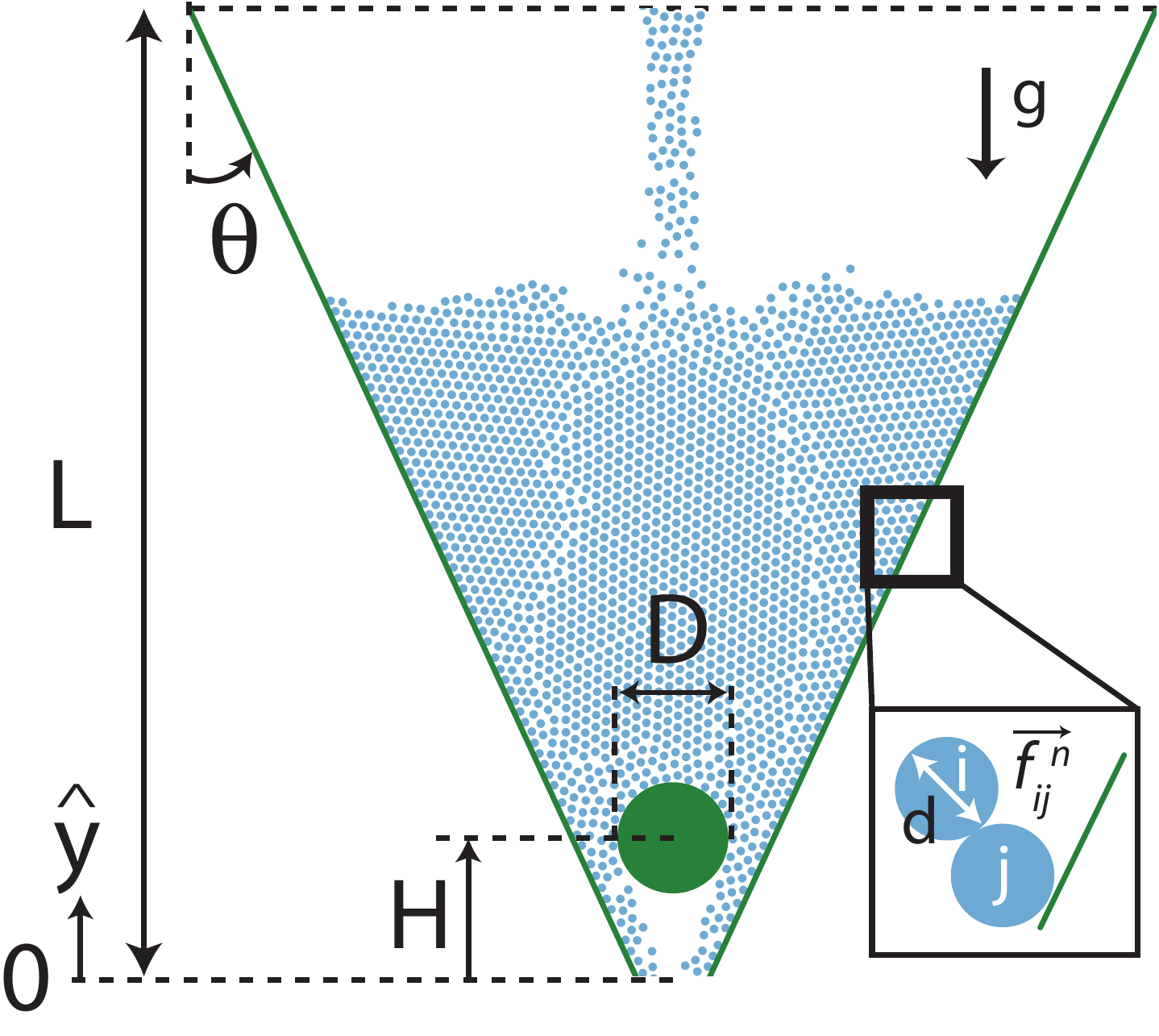}
\caption{\label{fig:model} (Color online) The MD simulation setup
  modeling steady gravity-driven hopper flow of frictionless discs
  (blue circles) of diameter $d$. An obstacle (green circle) of
  diameter $D$ sits at a height $H$ above the orifice of a symmetric
  hopper (green straight lines) with a height $L$ and a hopper angle
  $\theta$. Gravity $g$ is in the downward (-y) direction. The inset
  shows that the discs are subject to interparticle normal forces
  only.}
\end{figure}

In our MD simulations studying the gravity-driven discharging of
monodisperse or 50-50 bidisperse frictionless circular dry particles,
shown schematically in Fig.\ref{fig:model}, the hopper and the
obstacle have the same geometry as in the Tetris-like model. The
obstacle could be one disc of diameter $D$ or three
horizontally-aligned discs, each of diameter $D/3$ to resemble a flat
obstacle. The disc diameter $d$ of the monodisperse system is about
the same as the large disc diameter $d_l$ of the bidisperse system,
with $L/d=83$ and $L/d_l=82.857$. The size ratio between the obstacle
and a particle is $D/d=9.296$ and $D/d_l=9.28$. In the bidisperse
system, the diameter ratio between large and small discs is
$d_l/d_s=1.4$ to prevent artificial crystallization in a two
dimensional environment. There are $N$ discs in the system, where $N =
2048$ and $2712$ for the monodisperse and bidisperse systems,
respectively. These values ensure that the particles in each system
only fill the hopper up to about $2/3$ of its height while a steady
hopper flow is maintained. To maintain a constant number of particles
$N$ in our hopper flow simulation, a particle dropping out of the
hopper will reenter it from its top border by artificially shifting
the particle's vertical (y) position by a distance $L$ while keeping
its horizontal (x) position and velocities in both directions
unchanged.

\subsection{Interactions between objects within the system}
\label{Interactions}
In our MD simulation, initially orderly placed discs fall under
gravity and the system eventually reaches a steady state to form a
gravity-driven hopper flow. Each particle $i$ obeys Newton's
translational equation of motion
\begin{equation} \label{newton_law}
{{\mathord{\buildrel{\lower3pt\hbox{$\scriptscriptstyle\rightharpoonup$}}
      \over F} }_i} =
\mathord{\buildrel{\lower3pt\hbox{$\scriptscriptstyle\rightharpoonup$}}
  \over F} _i^{{\mathop{\rm int}} } +
\mathord{\buildrel{\lower3pt\hbox{$\scriptscriptstyle\rightharpoonup$}}
  \over F} _i^W +
\mathord{\buildrel{\lower3pt\hbox{$\scriptscriptstyle\rightharpoonup$}}
  \over F} _i^I +
\mathord{\buildrel{\lower3pt\hbox{$\scriptscriptstyle\rightharpoonup$}}
  \over F} _i^G =
        {m_i}{{\mathord{\buildrel{\lower3pt\hbox{$\scriptscriptstyle\rightharpoonup$}}
              \over a} }_i},
\end{equation}
where
${{\mathord{\buildrel{\lower3pt\hbox{$\scriptscriptstyle\rightharpoonup$}}
      \over F} }_i}$ is the total force acting on particle $i$ with
mass $m_i$, and acceleration
${{\mathord{\buildrel{\lower3pt\hbox{$\scriptscriptstyle\rightharpoonup$}}
      \over a}
  }_i}$. $\mathord{\buildrel{\lower3pt\hbox{$\scriptscriptstyle\rightharpoonup$}}
  \over F} _i^{{\mathop{\rm int}} }$,
$\mathord{\buildrel{\lower3pt\hbox{$\scriptscriptstyle\rightharpoonup$}}
  \over F} _i^W$,
$\mathord{\buildrel{\lower3pt\hbox{$\scriptscriptstyle\rightharpoonup$}}
  \over F} _i^I$ and
$\mathord{\buildrel{\lower3pt\hbox{$\scriptscriptstyle\rightharpoonup$}}
  \over F} _i^G$ are forces acting on particle $i$ from its contact
neighbors, the hopper wall, the obstacle, and gravity, respectively.

The simplest model of frictionless granular materials considers only
the interparticle normal forces \cite{gao09}. The interparticle force
$\mathord{\buildrel{\lower3pt\hbox{$\scriptscriptstyle\rightharpoonup$}}
  \over F} _i^{{\mathop{\rm int}} }$ on particle $i$ having $N_c$
contact neighbors can be expressed as
\begin{equation} \label{interparticle_force_law}
\mathord{\buildrel{\lower3pt\hbox{$\scriptscriptstyle\rightharpoonup$}}
  \over F} _i^{{\mathop{\rm int}} } = \sum\limits_{j \ne i}^{{N_c}}
        {[\mathord{\buildrel{\lower3pt\hbox{$\scriptscriptstyle\rightharpoonup$}}
              \over f} _{ij}^n} ({r_{ij}}) +
          \mathord{\buildrel{\lower3pt\hbox{$\scriptscriptstyle\rightharpoonup$}}
            \over f} _{ij}^{d}({r_{ij}})],
\end{equation}
where
$\mathord{\buildrel{\lower3pt\hbox{$\scriptscriptstyle\rightharpoonup$}}
  \over f} _{ij}^n({r_{ij}})$ and
$\mathord{\buildrel{\lower3pt\hbox{$\scriptscriptstyle\rightharpoonup$}}
  \over f} _{ij}^d({r_{ij}})$ are the interparticle normal force and
normal damping force defined below in Eqn.(\ref{particle_force}) and
Eqn.(\ref{particle_damping_force}), respectively.

Specifically, we assume that each frictionless particle $i$ is
subjected to a finite-range, purely repulsive linear spring normal
force from its contact neighbor $j$
\begin{equation} \label{particle_force}
\mathord{\buildrel{\lower3pt\hbox{$\scriptscriptstyle\rightharpoonup$}}
  \over f} _{ij}^n({r_{ij}}) = \frac{\epsilon}{{d_{ij}^2}}{\delta
  _{ij}}\Theta ({\delta _{ij}}){{\hat r}_{ij}},
\end{equation}
where $r_{ij}$ is the separation between disc particles $i$ and $j$,
$\epsilon$ is the characteristic elastic energy scale, $d_{ij} =
(d_i+d_j)/2$ is the average diameter, $\delta_{ij}=d_{ij}-r_{ij}$ is
the interparticle overlap, $\Theta(x)$ is the Heaviside step function,
and ${{\hat r}_{ij}}$ is the unit vector connecting particle centers.

Similarly, we consider only the interparticle normal damping force
proportional to the relative velocity between particles $i$ and $j$
\begin{equation} \label{particle_damping_force}
\mathord{\buildrel{\lower3pt\hbox{$\scriptscriptstyle\rightharpoonup$}} 
\over f} _{ij}^d({r_{ij}}) =  - b\Theta ({\delta _{ij}})({{\mathord{\buildrel{\lower3pt\hbox{$\scriptscriptstyle\rightharpoonup$}} 
\over v} }_{ij}} \cdot {{\hat r}_{ij}}){{\hat r}_{ij}},
\end{equation}
where $b$ is the damping parameter, and
${{\mathord{\buildrel{\lower3pt\hbox{$\scriptscriptstyle\rightharpoonup$}}
      \over v} }_{ij}}$ is the relative velocity between the two
particles. The normal damping force results in deduction of the
kinetic energy of the system after each pairwise collision.

The interaction force
$\mathord{\buildrel{\lower3pt\hbox{$\scriptscriptstyle\rightharpoonup$}}
  \over F} _i^W$ between particle $i$ and a hopper wall has an
analogous form to the interparticle interaction
$\mathord{\buildrel{\lower3pt\hbox{$\scriptscriptstyle\rightharpoonup$}}
  \over F} _i^{{\mathop{\rm int}} }$ with $\epsilon^W=2\epsilon$,
which means when a particle hits a wall, it experiences a repulsive
force as if it hit another mirrored self on the other side of the
wall. The particle-obstacle interaction force
$\mathord{\buildrel{\lower3pt\hbox{$\scriptscriptstyle\rightharpoonup$}}
  \over F} _i^I$ also has the same analogous form, and its value stays
zero if the hopper contains no obstacle. Finally,
$\mathord{\buildrel{\lower3pt\hbox{$\scriptscriptstyle\rightharpoonup$}}
  \over F} _i^G = - {m_i}g\hat y$, where $g$ is the gravitational
constant, and $\hat y$ is the unit vector in the upward
direction. There is no tangential interaction on particles in this
model, and therefore Newton's rotational equation of motion is
automatically satisfied.

The MD simulations in this study use the diameter $d$ and the mass $m$
of the monodisperse particles and the interparticle elastic potential
amplitude $\epsilon$ as the reference length, mass, and energy scales,
respectively. For the bidisperse system, the diameter $d_s$ and the
mass $m_s$ of the small particles separately replace $d$ and $m$. To
maintain a steady hopper flow without particles piling up to the upper
border of the hopper and bringing in unwanted boundary effects, we use
the dimensionless damping parameter to $b^*=db/\sqrt{m\epsilon}=0.5$,
the dimensionless gravity $g^*$ to $10^{-4}$, and a dimensionless time
step $dt^*=dt/{d}\sqrt {{m}/\epsilon}$ to $10^{-3}$ throughout this
study.

\subsection{Measuring the hopper flow rate}
\label{Measuring}
To measure the hopper flow rate while the obstacle is placed at a
given value of $H$ above the hopper orifice, we initiate one
simulation with orderly arranged particles. We also randomized size
identities for the bidisperse system. Then we wait for a time interval
$\Delta t^* = 5\times10^4$ until the system forgets the initial
arrangement and reaches a steady state to form a gravity-driven hopper
flow. After that, we count the number of particles passing the orifice
of the hopper within another $\Delta t^*$. For each value of $H$, we
use 18 different initial conditions to evaluate the average and the
variance of the actual flow rate $J_a$ in terms of number of particles
leaving the hopper per unit time. We define $J_o$ as the value of
$J_a$ while the hopper contains no obstacle.

\subsection{Simulation results}
\label{MD simulation results}
\begin{figure}
\includegraphics[width=0.45\textwidth]{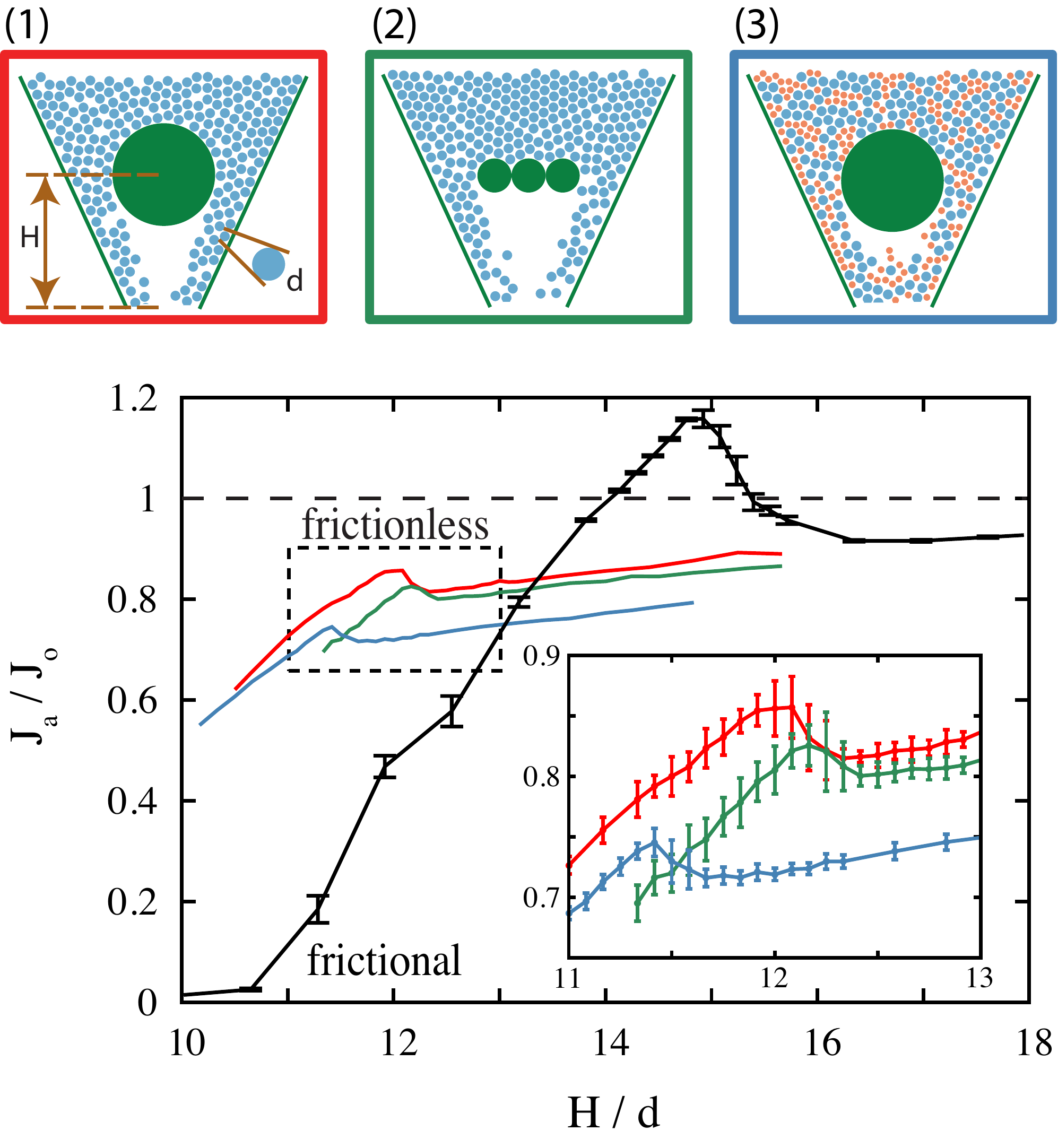}
\caption{\label{fig:hopper_flow_rate_comparison} (Color online)
  Averaged frictionless flow rates $J_a/J_o$ under different
  simulation setup: (1) monodisperse discs and a round obstacle (red);
  (2) monodisperse discs and a near flat obstacle (green); (3)
  bidisperse discs and a round obstacle (blue). The inset zooms in the
  dashed area. Each data point is obtained using 18 different initial
  conditions. A simulation snapshot of each frictionless setup is
  shown on the top with corresponding border color. The frictional
  flow rate of monodisperse discs and a round obstacle (black) is
  reproduced from Fig. 3(b) in Ref. \cite{alonso-marroquin12} for a
  quantitative comparison.}
\end{figure}

Our investigation contains two parts: A) To understand the influence
of the interparticle friction on the locally enhanced hopper flow
rate, we compare our frictionless results of the same hopper geometry
with the frictional data, copied from reference
\cite{alonso-marroquin12}, where monodisperse disc particles are
passing about a round obstacle. B) To understand the contribution of
the obstacle geometry or particle dispersity, we measured the flow
rates of frictionless discs in three cases: (1) monodisperse discs and
a round obstacle, (2) monodisperse discs and a flat obstacle, and (3)
50-50 bidisperse discs and a round obstacle. The results are shown in
Fig. \ref{fig:hopper_flow_rate_comparison}, where the actual flow rate
$J_a$, normalized by $J_o$, is plotted against the normalized obstacle
position $H/d$ or $H/d_l$ for the monodisperse or bidisperse
system. $J_o$ is $\approx 0.0319$ for the monodisperse system, and its
value increases by about $27\%$ to $\approx 0.0406$ for the bidisperse
system.

\subsubsection{Comparing with the frictional data}
\label{frictional_comparison}
Unlike their frictional counterparts, reproduced from reference
\cite{alonso-marroquin12}, frictionless particles start to flow
earlier and the normalized flow rate $J_a/J_o$ already reaches about
$60\%$ or higher as the obstacle is lifted to about ten-particles high
($H/d \approx 10$) above the hopper orifice. On the other hand, the
frictional normalized flow rate is only slightly above zero at a
similar $H/d$. The local peak value of frictional $J_a/J_o$ can be
greater than unity, while all three frictionless peaks have $J_a/J_o$
below unity with lower heights.

\subsubsection{Comparing between frictionless cases}
\label{frictional_comparison}
We find that the normalized hopper flow rate $J_a/J_o$ exhibits a
local peak in all three frictionless cases when the obstacle is lifted
to about eleven to twelve particles high ($H/d \approx 11$ to $12$)
above the orifice of the hopper. Among the three cases, the bidisperse
one with a larger $J_o$ exhibits its flow rate peak at $H/d \approx
11.4$, earlier than the other two monodisperse cases. Between the two
monodisperse cases with a round and a flat obstacle, the round
obstacle blocks the hopper flow less than the flat one, and the system
shows a peak slightly earlier at $H/d \approx 12$.

\subsubsection{An necessary condition for the local flow rate peak}
\label{Peak_N_condition}
\begin{figure}
\includegraphics[width=0.45\textwidth]{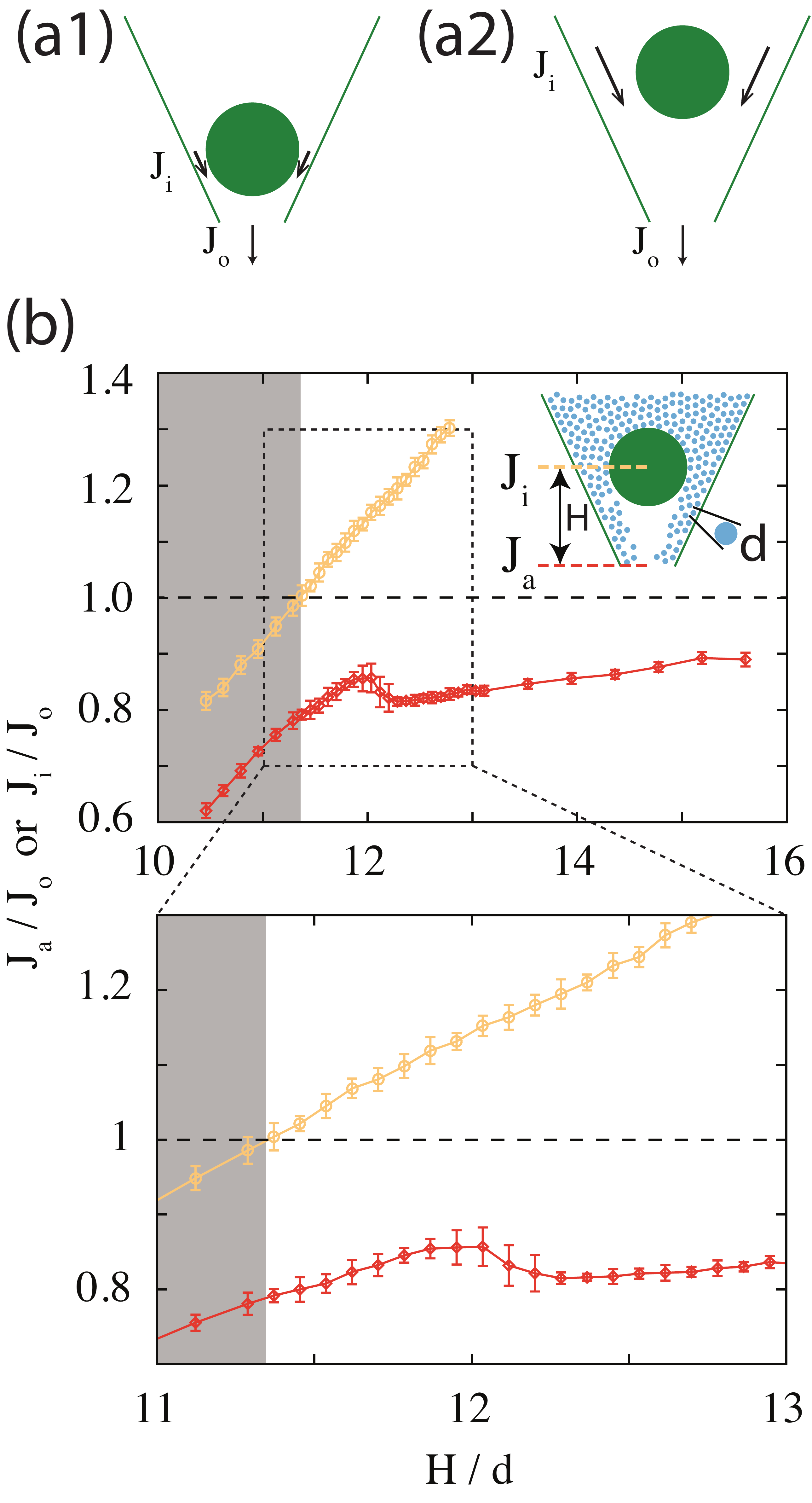}
\caption{\label{fig:N_cond_verification} (Color online) (a1) Schematic
  defining a fluidized flow regime where the flow rate $J_i$ at the
  obstacle is smaller than the maximum $J_o$ when the hopper contains
  no obstacle. (a2) Schematic defining a clogging flow regime where
  $J_i>J_o$. (b) Averaged flow rates, $J_i$ (orange) and $J_a$ (red),
  normalized by $J_o$ for the frictionless system with monodisperse
  disc particles and a round obstacle. Each data point is obtained
  using 18 different initial conditions. A zoomed-in plot at the
  bottom emphasizes the transition from $J_i < J_o$ (fluidized flow
  regime, shaded) to $J_i > J_o$ (clogging flow regime, unshaded),
  followed by the occurrence of a local peak of $J_a$.}
\end{figure}

Our results clearly show that none of the interparticle friction, the
obstacle geometry, or the particle dispersity is directly responsible
for the appearance of a local flow rate peak, though they do
effectively affect its position and magnitude. To better predict when
a flow rate peak occurs, we propose an indicator which is the flow
rate $J_i$, measured at the obstacle and normalized by $J_o$ while the
hopper contains no obstacle, as schematically shown in
Fig. \ref{fig:N_cond_verification}(a). Here we measure $J_i$ at the
same vertical height where the center of the obstacle is located, that
is, the height $H$ above the orifice of the hopper. Practically, we
measure $J_i$ by cutting off the part of the hopper below the center
of the obstacle so that the removed piece of hopper has no effect on
$J_i$.

When the obstacle is located closer to the hopper orifice, $J_i$ is
lower than $J_o$, defined as a fluidized flow regime as shown in
Fig. \ref{fig:N_cond_verification}(a1), and we should observe a
monotonic increase of the actual flow rate $J_a$. On the other
hand, when the obstacle is placed further away from the orifice, the
two internal passages between the obstacle and the two hopper walls on
its either side together can allow $J_i$ to become higher than $J_o$,
defined as a clogging flow regime, as shown in
Fig. \ref{fig:N_cond_verification}(a2). Presumably, $J_a$ can be
locally boosted in the clogging regime, due to a greater-than-unity
$J_i/J_o$ that cannot be smoothly constrained by the hopper until the
flow leaves its orifice, and therefore exhibits a local peak. $J_a$
then increases again as the position $H$ of the obstacle becomes
higher until it eventually reaches its maximum $J_o$. We believe that
$J_i/J_o>1$ is a necessary condition for observing a local flow rate
peak.

To offer simulation evidence showing the proposed necessary condition
is true, we plot $J_o$, $J_i$ and $J_a$ of the frictionless case of
monodisperse discs and a round obstacle as an example. The results are
shown in Fig. \ref{fig:N_cond_verification}(b). To numerically measure
$J_i$, we put particles dropping below $H$ back the top of the hopper
but slightly lower than its top border by a distance of
$0.1L$. Additionally, we place a lid with a dimensionless damping
parameter $b^*_l=50b^*$ at the top border of the hopper to prevent
fast-flying particles from escaping the simulation domain and conserve
the total number of particles $N$ in the system. As expected, we
observe a monotonic increase of $J_a$ while $J_i$ is below $J_o$. A
peak of $J_a$ occurs soon after $J_i/J_o>1$, and therefore we validate
the proposed necessary condition.

\bibliography{paper}

\end{document}